\documentclass[a4paper,twocolumn,superscriptaddress,11pt, accepted=2022-10-24]{quantumarticle}
\pdfoutput=1
\usepackage[numbers,sort&compress]{natbib}
\usepackage[utf8]{inputenc}
\usepackage[english]{babel}
\usepackage[T1]{fontenc}
\usepackage{nicefrac}
\usepackage{graphicx}% Include figure files
\usepackage{bm}% bold math
\usepackage{amssymb}
\usepackage{color}
\usepackage{amsmath, mathrsfs}
\usepackage{amstext}
\usepackage{float}
\usepackage[outdir=./]{epstopdf}
\usepackage{latexsym}
\usepackage[usenames,dvipsnames]{xcolor}
\usepackage[colorlinks=true,citecolor=Blue,linkcolor=RubineRed,urlcolor=Blue]{hyperref}
\usepackage{bbold}% bold math
\usepackage{cleveref}
\DeclareFontFamily{OT1}{pzc}{}
\DeclareFontShape{OT1}{pzc}{m}{it}{<-> s * [0.900] pzcmi7t}{}
\DeclareMathAlphabet{\mathpzc}{OT1}{pzc}{m}{it}

\AtBeginDocument{%
    \newwrite\bibnotes
    \def\bibnotesext{Notes.bib}
    \immediate\openout\bibnotes=\jobname\bibnotesext
    \immediate\write\bibnotes{@CONTROL{REVTEX41Control}}
    \immediate\write\bibnotes{@CONTROL{%
    apsrev41Control,author="08",editor="1",pages="1",title="0",year="1"}}
     \if@filesw
     \immediate\write\@auxout{\string\citation{apsrev41Control}}%
    \fi
}%

\def\dg{^\dagger}
\newcommand{\beq}{\begin{equation}}
	\newcommand{\eeq}{\end{equation}}
\newcommand{\beqa}{\begin{eqnarray}}
	\newcommand{\eeqa}{\end{eqnarray}}

\newcommand{\bal}{\begin{align}}
	\newcommand{\eal}{\end{align}}

\newcommand{\ket}[1] {\vert #1 \rangle}

\newcommand{\braket}[2] {\langle #1 | #2 \rangle}
\newcommand{\Tr}{\mathrm{Tr}}

\newcommand{\pablo}{\color{black}}

\DeclareMathOperator*{\Max}{max}

\newcommand{\av}[1]{\left \langle #1 \right \rangle}

\newcommand{\avb}[1]{\langle #1 \rangle_\beta}

\newcommand{\be}[1]{\mathbf{#1}}
\newcommand{\mc}[1]{\mathcal{#1}}
\newcommand{\mr}[1]{\mathrm{#1}}

\newcommand{\ha}[1]{\hat{#1}}
\newcommand{\ti}[1]{\tilde{#1}}

\newcommand{\mbb}[1]{\mathbb{#1}}

\begin{document}
	
	\title{Analyticity constraints  bound\newline the decay of the spectral form factor}
	\author{Pablo Martinez-Azcona \href{https://orcid.org/0000-0002-9553-2610}{\includegraphics[scale=0.05]{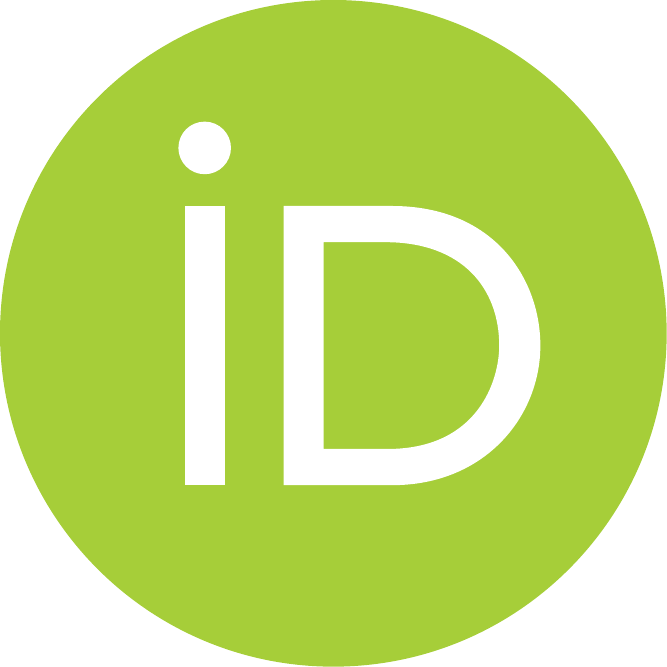}}}
	\affiliation{Department of Physics and Materials Science, University of Luxembourg, L-1511 Luxembourg}
\author{Aur\'elia Chenu \href{https://orcid.org/0000-0002-4461-8289}{\includegraphics[scale=0.05]{orcidid.pdf}}}
	\affiliation{Department of Physics and Materials Science, University of Luxembourg, L-1511 Luxembourg}

\maketitle

	%\date{\today}

		\begin{abstract}  
		
		Quantum chaos cannot develop faster than $\lambda \leq 2 \pi/(\hbar \beta)$ for systems in thermal equilibrium [Maldacena, Shenker \& Stanford, JHEP (2016)]. This `MSS bound' on the Lyapunov exponent $\lambda$ is set by the width of the strip on which the regularized out-of-time-order correlator is analytic. 
		{\pablo We show that similar constraints also bound the decay of the spectral form factor (SFF), that measures spectral correlation and is defined from   
		%We show that similar analyticity constraints also bound the evolution of other dynamical quantities.  
  %We first find a family of functions that admit a universal bound inspired by the MSS bound, and then detail the case of the spectral form factor (SFF), 
 the Fourier transform of the two-level correlation function. %---it can also be understood as the survival probability of the coherent Gibbs state.		
Specifically, the \textit{inflection exponent} $\eta$, that we introduce to characterize the early-time decay of the SFF, is bounded as $\eta\leq \pi/(2\hbar\beta)$. 
		This bound is universal and exists outside of the chaotic regime.} The results are illustrated in systems with regular, chaotic, and tunable dynamics, namely the single-particle harmonic oscillator, the many-particle Calogero-Sutherland model, an {\pablo ensemble} from random matrix theory, and the quantum kicked top. 
		{\pablo The relation of the derived bound with other known bounds, including quantum speed limits, is discussed.  }

	\end{abstract}

	\section{Introduction}
	How fast can a given quantum system evolve? Bounds setting limits on the evolution of dynamical quantities have  proven to be useful tools and brought a great deal of insight. 
	On the one hand, quantum speed limits, that determine the minimum time for evolution under quantum dynamics \cite{mandelstam_uncertainty_1991, margolus_maximum_1998, levitin_fundamental_2009, del_campo_quantum_2013, taddei_quantum_2013},  have been the focus of intense studies in both the quantum \cite{ pfeifer_generalized_1995, muga_time_2008, muga2009time, frey_quantum_2016, deffner_quantum_2017} and classical   \cite{shanahan_quantum_2018, okuyama_quantum_2018, poggi_diverging_2021} realms, with a recent unification of the two realms for speed limits on observables  \cite{garcia-pintos2022}.  Beyond their fundamental relevance, these bounds  have become useful tools in the study of quantum information and technologies \cite{bekenstein_energy_1981,lloyd_ultimate_2000, del_campo_scrambling_2017}, many-body physics \cite{bukov2019,fogarty2020,delcampo2021}, and find applications in quantum control \cite{caneva_optimal_2009, funo2017, deffner_quantum_2017} and quantum metrology \cite{giovannetti_advances_2011,beau2017}.

	On another hand, a universal bound on quantum chaotic dynamics has been recently proposed \cite{maldacena_bound_2016}. 
		It sets a limit on 
	the quantum Lyapunov exponent $\lambda$, defined from the `Out-of-Time-Order Correlator' (OTOC). This correlator was originally proposed in the context of superconductivity \cite{larkin_quasiclassical_1969} and has been extended to the high energy \cite{hashimoto_out--time-order_2017, hanada_universality_2018,gharibyan_quantum_2019, akutagawa_out--time-order_2020, kobrin_many-body_2021} and quantum information 	\cite{rozenbaum_lyapunov_2017, shen_out--time-order_2017, tsuji_out--time-order_2018,sieberer_digital_2019, fortes_gauging_2019, chavez-carlos_quantum_2019, keles_scrambling_2019, lewis-swan_unifying_2019, pg_out--time-ordered_2021, pilatowsky-cameo_positive_2020, wang_microscope_2021-1, yin_quantum_2021}
	 communities. In the semiclassical limit and for a certain time range, the OTOC behaves exponentially and defines a proper analog of the Lyapunov exponent \cite{kitaev_hidden_2014}. Maldacena, Shenker and Stanford \cite{maldacena_bound_2016} conjectured that, for any thermal state, this exponent is bounded as $\lambda \leq 2 \pi/ (\hbar \beta)$. This finding motivated considerable attention within the community \cite{kurchan_quantum_2018, tsuji_bound_2018, turiaci_inelastic_2019, murthy_bounds_2019,kundu_subleading_2022,  kobrin_many-body_2021, pappalardi_low_2022}. 
	 {\pablo The  original derivation relied  on  the analytic continuation of the regularized OTOC to complex times $t + i \tau$ and the region in which it is analytic, and has been proven alternatively since then  \cite{tsuji_bound_2018}. The bound itself has been extended beyond quantum chaos and linked to the fluctuation-dissipation theorem \cite{pappalardi2022, tsuji_out--time-order_2018}.  %\reminder{Cite \cite{grozdanov2021} here}. 
	 Similar arguments based on analyticity have yielded bounds on other dynamical quantities \cite{grozdanov_bounds_2021} or  been used to characterize dynamical phase transitions \cite{heyl_dynamical_2013}. }

	%
	 %To the best of our knowledge, it has not yet been asked  if such property sets universal bounds on other dynamical quantities, or if such bounds are unique to chaotic systems. 

	 Here, we show that the mathematical property used to derive the MSS bound applies to  quantities other than the OTOC, and sets bounds that are not restricted to chaotic behavior.
	{\pablo  Particularly, we find these bounds on a dynamical quantity that measures spectral correlation and is very widespread in the quantum chaos community, the spectral form factor \cite{barbon_very_2003, barbon_long_2004, papadodimas_local_2015, cotler_black_2017, cotler_chaos_2017, del_campo_scrambling_2017}. }

	In Section \ref{sec:BoundSFF}, we show that the region of analyticity also imposes a universal  bound on its {\pablo early-time} evolution, that  holds for any system
	 ranging from regular to chaotic behavior, and can be very tight. 
	 We thus {\pablo also} extend such universal bounds beyond the context of quantum chaos, {\pablo and propose an interpretation of the quantity we bound in terms of the average energy at complex  temperature.}
	 Section~\ref{sec:examples} illustrates our findings in systems that are conceptually very different and representatives of regular and chaotic dynamics: the harmonic oscillator, the Calogero-Sutherland model, the Gaussian Unitary Ensemble from random matrix theory \cite{mehta_random_2004} and the quantum kicked top \cite{haake_classical_1987}, {\pablo which is one of the earliest model introduced to study chaos and can mimic the behavior of any members of a universality class displayed by random matrix theory}. In Section~\ref{sec:QSL}, we show how the derived bound compares to known bounds, starting with quantum speed limits.

	%%%%%%%%%%%%%%%%%%%%%%%%%%%%%%%%%
		\begin{figure*}
		\centering
		\begin{tikzpicture}[scale=1]
			\fill [blue!40] (-1.67, -0.5) rectangle (1.67, 0.5);
			\draw[black, thin, ->] (-1.67, 0) -- (1.67, 0) node[anchor=west] {$t$};
			\draw[black, thin, ->] (0, -1) -- (0, 1) node[anchor=west] {$\tau$};
			\draw[black, very thick] (-1.67, 0.5) node[anchor=south west] {$\tau =\beta \hbar$} -- (1.67, 0.5);
			\draw[black, very thick] (-1.67, -0.5) node[anchor=north west] {$\tau =-\beta \hbar$} -- (1.67, -0.5);
			%\draw[black, very thick] (0, -0.5)-- (0, 0.5); 
			\draw[black, very thick, ->] (2.1, 0) -- (4.8, 0) node[anchor=north east] {$z = \frac{e^{\frac{\pi}{2 \beta \hbar}(t + i \tau)}-1}{e^{\frac{\pi}{2 \beta \hbar}(t + i \tau)}+1}$} node[anchor = south east]{conformal map}; 
			\filldraw[color=black, fill=blue!40, very thick] (6, 0) circle (0.5);
			\draw[black, thin, ->] (5, 0) -- (7.1, 0 ) node[anchor=west] {$\mr{Re}(z)$};
			\draw[black, thin, ->] (6, -1) -- (6, 1) node[anchor=west] {$\mr{Im}(z)$} node[anchor=north east] {$1$}  ;
			\draw[red, very thick,->,shorten >=1pt] (6.4, 0.5) to [out=45,in=315,loop,looseness=2.5] (6.4, -0.5)  node[anchor=north] {$f_z\:\:$};
			\draw[black, very thick, ->] (8.4, 0) -- (11.5, 0) node[midway,above] {Schwarz-Pick} node[midway,below] {$f_{t+ i \tau}=\tilde{S}_{\beta, t + i \tau}$} node[anchor=west] {$\boxed{\: \eta \leq \frac{\pi}{2 \beta \hbar}\: }$};
		\end{tikzpicture}
		\label{fig:proof}
		\caption{{\pablo \textbf{Illustration of the proof that bounds the inflection exponent $\eta$.} From left to right: The strip ($t,\tau$) of the complex plane in which $f_{t+i\tau}$ is analytic is mapped to the unit disk using the conformal map \eqref{map}. For an analytic function $f_z$ that maps $z$ from the unit disk to $f_z$ onto the same unit disk and fulfills $|f_z|\leq1$,  the Schwarz-Pick theorem  yields a bound on $f_z$ \eqref{SPTheorem}. For the spectral form factor, this results in a bound on the inflection exponent that is fixed by the system temperature and the Planck constant, see Eq. \eqref{bound_eta}. }
		%This setting gives another example, apart from MSS bound, where the width of the strip in which a quantity is analytic yields a universal bound on a related quantity. 
		\label{fig:proof}}
	\end{figure*}
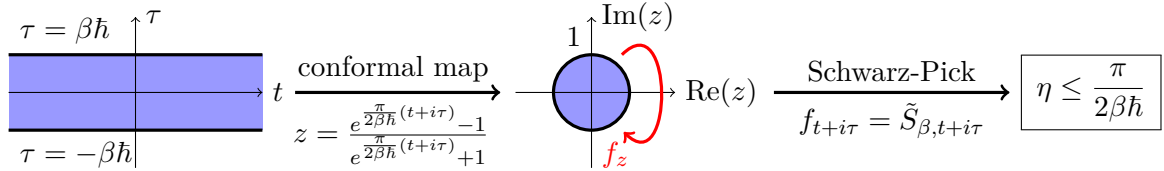
	
	{\pablo
	\section{A bound on the decay of the spectral form factor}  
	\label{sec:BoundSFF}
	%\subsection{A universal bound on exponential decay \label{sec:ProofBound}}
	In this Section, we find that the spectral form factor, after the short-time Gaussian decay, decays exponentially with an exponent bounded by the temperature and Planck constant only. The mathematical proof follows that used by MSS \cite{maldacena_bound_2016} and extends its range of applicability to  short times, with no neglected terms.  
	
	\subsection{Early-time decay of the SFF}
	%We consider exponential decay and show that the exponent can be bounded by analyticity constraints. The proof is inspired by the mathematical property used to derive the MSS bound on the Lyapunov exponent, $\lambda \leq 2 \pi/\hbar \beta$, that we briefly recast in the App. \ref{sec:MSSbound} for completeness. 
	
	}
	
	The spectral form factor (SFF) is an efficient  tool for determining the spectral properties of a system, and it is the simplest nontrivial measure of spectral correlations \cite{Bertini2018}. This dynamical quantity is the Fourier transform of the two-level correlation function and can be interpreted as the fidelity between a coherent Gibbs state \cite{del_campo_scrambling_2017, xu_extreme_2019, del_campo_decoherence_2020, xu_thermofield_2021, cornelius_spectral_2021}, $\ket{\psi_\beta}=Z_\beta^{-1/2}\sum_n e^{-\beta E_n/2} \ket{n}$, and its time evolution, namely
	\begin{align}\label{SFF}
		S_{\beta, t}&= |\braket{\psi_\beta}{e^{-\frac{i}{\hbar} \hat{H} t}|\psi_\beta} |^2= \left|\frac{Z_{\beta +i t/\hbar}}{Z_\beta}\right|^2,\\ \notag
		&=\frac 1{Z_\beta^2}\sum_{m,n}e^{-(\beta + \frac{i}{\hbar} t)E_n}e^{-(\beta - \frac{i}{\hbar} t)E_m}.
	\end{align}
	It reads as the normalized analytical continuation of the partition function with $Z_\beta = \Tr (e^{-\beta \hat{H}})= \sum_n e^{-\beta E_n}$, $E_n$ being the system eigenenergies and $\beta$ the inverse temperature. The SFF decays from its initial unity value with a Gaussian shape at short times \cite{del_campo_scrambling_2017}. For systems with correlated eigenenergies such as chaotic ones, the SFF reaches a dip and then goes up with a ramp, interpreted as a signature of chaos \cite{cotler_black_2017, cotler_chaos_2017, del_campo_scrambling_2017}, and plateaus at a constant value, fixed by the dimension of the Hilbert space $N$ and the  temperature. 
	
		While the SFF is widely used in chaos because of this characteristic shape, we  are here interested in {\pablo its early-time decay, that is, after the initial Gaussian decay and before the onset of chaotic features.} We are thus not restricting ourselves to any dynamical regime. Specifically, we consider the time  $t_0$  at which $\ln(S_{\beta, t})$ has a first inflection point---its second derivative vanishes. 
	{\pablo	To characterize the decay around this time, we introduce the \textit{inflection exponent} 
		%Around $t_0$ the SFF shows an exponential decay, that we characterized by introducing the \textit{inflection exponent} $\eta$ defined as
	\begin{equation}\label{def_eta}
		\eta %= \Max_t\left|\frac{\dot{S}_{\beta, t}}{S_{\beta, t}}\right|
		=\left|\frac{\dot{S}_{\beta, t_0}}{S_{\beta, t_0}}\right|,
	\end{equation}
	that corresponds to $\Max_t\big|\dot{S}_{\beta, t}/ S_{\beta, t}\big|$. % and depends on the inverse temperature.
	%
	%This exponent depends on the inverse temperature $\beta = 1/k_B T$ and  corresponds to $\Max_t\big|\dot{S}_{\beta, t}/ S_{\beta, t}\big|$.
	%
	Around this time maximum, the function can be approximated by a constant up to first order in time, \textit{i.e.} $|\dot{S}_{\beta, t}/S_{\beta, t}|= \eta + \mc O((t - t_0)^2)$. 
	%This yields a differential equation which gives the approximate behavior 
	So for $t$ close to $t_0$, the SFF decays exponentially $S_{\beta, t} \sim S_{\beta}^{0} e^{-\eta t}$, with $S_{\beta,0}$ a constant. 
	We find that this early-time decay of the SFF is actually bounded. 	
	}
	\\ 
{\pablo 
	\subsection{Bound on the SFF exponential decay}

	%The main ingredient for the proof is the \textit{Schwarz-Pick theorem}. It states that if a function of a complex variable $f_z: z \in \mr D \rightarrow \mr D$ maps the unit disk $\mr D$ into itself and $f$ is analytic, the following inequality holds
	The proof that $\eta$ is bounded is sketched in Fig.~\ref{fig:proof}. 
	We rely on the analytical continuation of the SFF to complex times, that we define as
	\begin{equation} \label{analyticalSFF}
		S_{\beta, t + i \tau}=\frac{Z_{\beta-\tau/\hbar+it/\hbar}Z_{\beta + \tau/\hbar -i t/\hbar}}{Z_{\beta}^2},
	\end{equation}
and  tools from complex analysis. Specifically, we use  the \textit{Schwarz-Pick theorem}: For an analytic
%\footnote{An analytic function of a complex variable $f_z$ can be written as a  power series convergent at every point of its domain. 
% All complex analytic functions are holomorphic and vice versa. A holomorphic function $f_z$ is complex differentiable at every point of its domain.}
%\footnote{An analytic function of a complex variable $f_z$ can be written as a  power series convergent at every point of its domain.}
\footnote{A function of a complex variable $f_z$ is analytic in a region $ \mc R$ if it can be written as a  power series convergent at every point of $\mc R$.}
  function $f_z$ of a complex variable $z = x+ i y$ that maps the unit disk $\mr D$ into itself $f_z : z \in \mr{D} \rightarrow \mr D$, the following inequality holds
	\begin{eqnarray}\label{SPTheorem}
		\frac{|\mr d f_z|}{1 - |f_z|^2}\leq \frac{| \mr d z|}{1-|z|^2},
	\end{eqnarray}
	with the differentials $\mr df_z = \frac{\partial f_z}{\partial x} \mr dx  + \frac{\partial f_z}{\partial y} \mr dy$ and $\mr dz = \mr dx + i \mr dy$. Note that the condition of mapping the unit disk into itself is equivalent to satisfying $|f_z|\leq 1 \; \forall  \; z \in \mr D$.

	The spectral form factor at complex time, $t\rightarrow t + i \tau$ see Eq. \eqref{analyticalSFF}, is  analytic on the strip of the complex plane $-\beta \hbar \leq \tau \leq +\beta \hbar$ and $t \in \mbb R$, as shown in App. \ref{app:AnalytSFF}. This strip holds for systems with a Hilbert space of infinite dimension with unbounded energies, as well as for systems with a finite-dimensional Hilbert space---in this latter case, the SFF is analytic on the whole complex plane\footnote{As is also stated in \cite{heyl_dynamical_2013} for the analytical continuation of the partition function, we find that the SFF is an entire function (analytic in the whole complex plane)  of $z$ for a finite dimensional Hilbert space.}
	 because it is given by a finite linear combination of complex exponentials, analytic on the whole complex plane.  
	 In contrast to quantities like the OTOC or two-point correlation functions, this quantity is also analytic at $t=0$, as we verify in App. \ref{app:AnalytSFF}. 
	%Our interest relies on the Spectral Form Factor. When time is complexified $t \rightarrow t+ i \tau$ this quantity $S_{\beta, t+ i \tau}$ is always analytic on a stripe of the complex plane, $-\beta \hbar \leq \tau \leq +\beta \hbar$. In contrast to quantities like the OTOC or 2-point correlation functions, this quantity is also analytic at $t=0$, see App. \ref{app:AnalytSFF} for a formal proof.  There is a subtlety here, if the Hilbert space is finite dimensional, since the SFF is given by a finite linear combination of analytic functions, it will be analytic in the whole complex plane, and therefore also analytic on the stripe of interest. However when we consider the case of a infinite dimensional Hilbert space, with unbounded energies, the SFF is analytic only on this stripe of the complex plane. 
	
	The region in which the SFF is analytic  can be mapped to the unit disk using  
	\begin{equation}\label{map}
	z = \frac{e^{\frac{\pi}{2 \beta \hbar}(t + i \tau)}-1}{e^{\frac{\pi}{2 \beta \hbar}(t + i \tau)}+1}.
\end{equation}
	Using inequality \eqref{SPTheorem} along the real line $\tau =0$ yields 
	\begin{eqnarray}
		\frac{1}{1 - |f_t|^2}\left|\frac{\mr d f_t}{\mr d t}\right|\leq \frac{1}{2}  \frac{\pi}{2 \beta \hbar},
	\end{eqnarray}
	which, assuming $f_t \in \mbb R$ in the real line $t$, can be recasted as
	\begin{eqnarray}\label{ineq1over1mf}
		\frac{1}{1 - f_t}\left|\frac{\mr d f_t}{\mr d t}\right|\leq \frac{1+f_t}{2}  \frac{\pi}{2 \beta \hbar}\leq   \frac{\pi}{2 \beta \hbar}.
	\end{eqnarray}
	This inequality is very similar to the one used by MSS \cite{maldacena_bound_2016} but with one difference, which is key to derive bounds at short times: the terms  $\mc O(e^{-4 \pi t/(\beta \hbar) })$ are not neglected. These terms could be important at early times and therefore analyticity at $t=0$ plays an important role.
	
	%It is easy to verify that $f_z = 1- S_{\beta,t+i\tau}$ fulfills $|f_z|\leq 1$ 
	Importantly, we need $|f_z|\leq 1$ in all points of the domain to obtain the inequality \eqref{SPTheorem} from the Schwarz-Pick theorem. To do so, we define the \textit{modified spectral form factor}, 
\begin{equation} \label{analyticalSFFnorm}
	\tilde{S}_{\beta, t + i \tau}=1-\frac{Z_{\beta-\frac {\tau} {\hbar}+\frac i {\hbar}t}Z_{\beta+\frac {\tau} {\hbar}-\frac i {\hbar}t}}{Z_{\beta-\frac{\tau}{\hbar}}Z_{\beta+\frac{\tau}{\hbar}}}
\end{equation}
that we introduce such that $\tilde{S}_{\beta, t} = 1- S_{\beta,t}$ on the real line and, since $|Z_{\beta-\nicefrac {\tau} {\hbar}+\nicefrac {it} {\hbar}}Z_{\beta+\nicefrac {\tau} {\hbar}-\nicefrac {it} {\hbar}}|\leq  Z_{\beta-\nicefrac{\tau}{\hbar}}Z_{\beta+\nicefrac{\tau}{\hbar}}$, one expects $|\tilde{S}_{\beta, t+ i \tau}|\leq 1$---there is however a subtlety  that we discuss in the next subsection.  Note that this function still preserves analyticity in the same domain, since the denominator is an analytic function that is never zero and the numerator is analytic on the strip (as shown in App. \ref{app:AnalytSFF}). This function has the same inflection exponent that the SFF and fulfills the conditions to apply the Schwarz-Pick theorem. 
Choosing $f_z$ to be the modified  SFF, $f_{t+i \tau} = \ti S_{\beta, t+ i \tau}$, the l.h.s. of  inequality \eqref{ineq1over1mf} simplifies and we obtain a bound on the inflection exponent: 
\begin{equation}\label{bound_eta}
	\eta \leq \frac{\pi}{2 \beta \hbar}.
\end{equation}
}This is our main result. It means that around the inflection time $t_0$, the fastest possible decay of the SFF is proportional to the temperature of the system.

		\begin{figure*}
	\includegraphics[width = \linewidth]{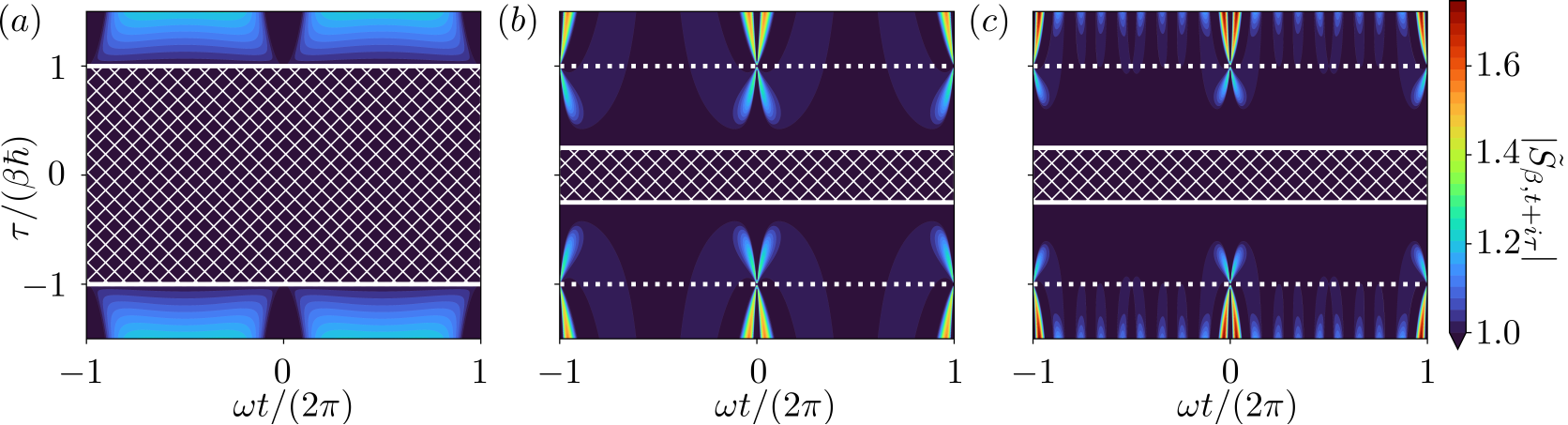}
	\caption{{\pablo \textbf{Modulus of the modified Spectral Form Factor for complex time $|\tilde{S}_{\beta, t+ i \tau}|$} for (a) a single harmonic oscillator, (b) a set of independent harmonic oscillators with partition function $Z_{\beta}=(Z_\beta^\textsc{ho})^N$ and (c) the Calogero-Sutherland model.  The scale has been chosen to highlight the regions in which $|\tilde{S}_{\beta, t+ i \tau}|>1$. For both (b) and (c) $N=4$ particles were considered. The dotted white lines represent the standard strip $-\beta \hbar \leq \tau \leq \beta \hbar$ and the solid lines with the hatch represent the renormalized strip $-\beta \hbar/N \leq \tau \leq \beta \hbar/N$. The function is bounded in this renormalized strip, $|\tilde{S}_{\beta, t+ i \tau}|\leq 1$, where the Schwarz-Pick theorem holds. }
	\label{fig:modSFF_complext}}
\end{figure*}

{\pablo	
	\subsection{Bound for an extensive inflection exponent}
	
	Note that Eq. \eqref{bound_eta} holds provided that $\eta$ is intensive. However, this need not always be the case since,  in the most general case, the partition function $Z_\beta=e^{-\beta F}$ is defined from   the Helmholtz free energy $F$, which can be  extensive. 
	 Since $\eta$ represents the exponential decay of the SFF that is defined from the partition function, it is expected to be extensive whenever $F$ is extensive.  Whenever $\eta$ is extensive is some quantity $d$ (\textit{e.g.} the dimension of the system, the number of particles, or the central charge in conformal field theory), we conjecture a correction of the  bound as 
	\begin{equation}\label{etaD}
		\frac{\eta}{d} \leq \frac{\pi}{2 \beta \hbar}.
	\end{equation}
	The idea behind this correction is that   the modulus of the modified SFF $|\tilde{S}_{\beta, t+ i \tau}|$ can be bigger than unity in the strip $-\beta \hbar \leq \tau \leq \beta \hbar$ for the models in which $\eta$ is extensive. A way to recover the condition that $f_z$ maps the unit disk into itself is to shrink the strip to $-\beta \hbar/d \leq \tau \leq \beta \hbar/d$.  % in which $|\tilde{S}_{\beta, t+ i \tau}|\leq 1$. 
	Substituting $\beta \rightarrow \beta/d$ in \eqref{bound_eta} yields the bound for an extensive inflection exponent given in Eq. \eqref{etaD}.

	Figure \ref{fig:modSFF_complext} illustrates the correction of the strip for $N$  independent harmonic oscillators and for the Calogero Sutherland model.  	
	%One can always shrink the strip to a smaller strip in which $|\tilde{S}_{\beta, t+ i \tau}|\leq 1$, however we think (conjecture?) that this is related to the cases where $\eta$ is extensive.  We illustrate this idea in Fig. \ref{fig:modSFF_complext} for a set of $N$ independent Harmonic Oscillators (b) and for the Calogero-Sutherland model (c). 
	In both cases, $|\ti S_{\beta, t+i \tau}|$ is bigger than one for some regions of the original strip but not within the shrunk strip, so the Schwarz-Pick theorem applies. %, rendering the conjectured bound for extensive $\eta$ \eqref{etaD}. 
	This conjecture is further detailed in Sec. \ref{sec:CS} for the Calogero-Sutherland model, that is extensive in the number of particles. 
	\subsection{An interpretation of the inflection exponent}
	
The spectral form factor \eqref{SFF} can be written as the expectation value of the evolution operator $\hat{U}_t = e^{- i \hat{H} t / \hbar}$ for the thermal state $\hat{\rho}_\beta=e^{-\beta \ha H}/Z_\beta$, that is, $S_{\beta, t} = \avb{\ha U_t} \avb{\ha U_t\dg}$ where $\avb{\bullet}=\Tr(\bullet \ha \rho_\beta)$. Thus, 
\begin{equation}\label{SdotS}
	\frac{\dot S_{\beta, t}}{S_{\beta, t}}= - \frac{i}{\hbar} \frac{\avb{\ha H \ha U_t}}{\avb{\ha U_t}} + \frac{i}{\hbar} \frac{\avb{\ha H \ha U_t\dg}}{\avb{\ha U_t\dg}}.
\end{equation}
The inflection exponent, as defined in \eqref{def_eta}, corresponds to this expression evaluated at the inflection time $t_0$, \textit{i.e.} when the function has its first minimum. Since the two terms in \eqref{SdotS} are complex conjugates of each other, the inflection exponent $\eta$ can be recast as
\begin{equation}\label{interpretation_eta}
	\eta = \frac{2}{\hbar} \mr{Im}\left( \frac{\avb{\ha H \ha U_{t_0}\dg}}{\avb{\ha U_{t_0}\dg}} \right)=  \frac{2}{\hbar} \mr{Im}\langle{\ha H}\rangle_{\beta - \frac{i}{\hbar}t_0} ,
\end{equation}
which corresponds to the imaginary part of the average energy at complex $\beta$, where the complex part of the inverse temperature is fixed by the inflection time $t_0$. 
Note that, when dealing with chaotic systems defined over an ensemble,  the value of the inflection time $t_0$ cannot be determined from a single realization but rather from the averaged SFF. This is because the SFF is not self-averaging \cite{prange_spectral_1997}, as we detail in the examples below. In this case, the above definition \eqref{interpretation_eta} is not computationally efficient unless exact analytic forms of $\avb{\ha H}$ are known.

Finally, the derived bound poses the maximum value 
\begin{equation}
 \mr{Im}\left( \frac{\avb{\ha H \ha U_{t_0}\dg}}{\avb{\ha U_{t_0}\dg}} \right)=  \mr{Im}\langle{\ha H}\rangle_{\beta - \frac{i}{\hbar}t_0} \leq \frac{\pi}{4 \beta}.
\end{equation}
This interpretation of the inflection exponent highlights that the initial Gaussian decay, determined by $\Delta \ha H^2$ \cite{del_campo_scrambling_2017}, is followed by the early-time exponential decay with a different exponent, which is determined by the imaginary part of the expectation value of $\ha H$ at complex $\beta$. 
}

%}		

	%

	%%%%%%%%%%%%%%%%%%%%%%%%%%

	\section{Examples \label{sec:examples}}
		In order to  illustrate the derived bound in some specific setups, we choose four conceptually very different systems, that respectively exhibit regular (both single- and many-particle), chaotic, and tunable (between regular and chaotic) dynamics. Namely, we compute the SFF and look at the inflection exponent in the harmonic oscillator, the Calogero-Sutherland model, an ensemble from random matrix theory, and the quantum kicked top.

	\subsection{Integrable system: the harmonic oscillator}  
\begin{figure}[h]
			\centering
			\includegraphics[width=0.85\columnwidth]{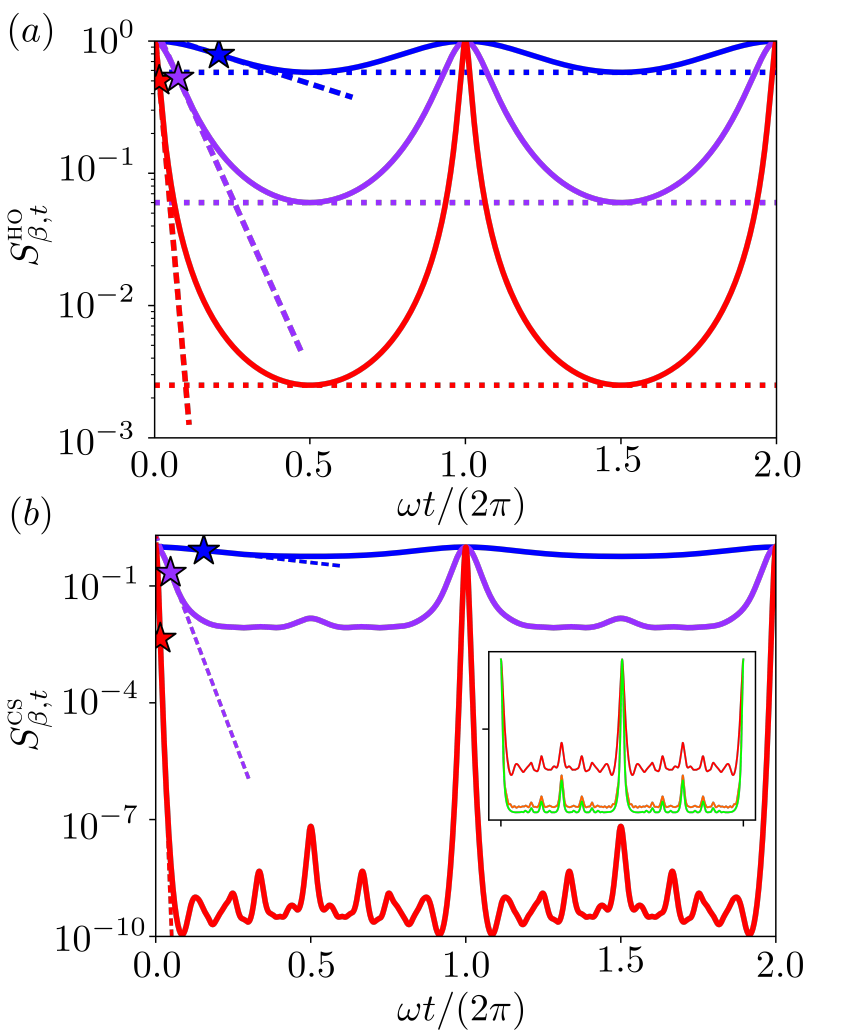}
			\caption{\textbf{Time evolution of the spectral form factor} for the (a)  harmonic oscillator \eqref{S_HO}  and  (b)  Calogero-Sutherland model \eqref{S_CS} at $ \beta \hbar \omega=2$ (blue), $0.5$ (purple) and $0.1$ (red). The dashed lines represent the function $e^{-\eta t}$ around the inflection time $t_0$, marked by a star. (a)   The dotted lines mark the SFF minimum values. (b) SFF for $N = 10$. The inset shows the SFF evolution at $\beta \hbar \omega = 0.1$ for $N=10$ (red), $30$ (dashed orange) and $100$ (solid green) interacting particles---the two red curves coincide. {\pablo Increasing  the temperature or the number of particles makes the SFF reach lower values.}}
			\label{fig:SFF-integrable}
		\end{figure}

We start with a single particle in an harmonic trap, which Hamiltonian
	\beq
	\ha H= \hbar \omega \left(\ha a^\dagger \ha a + \frac{1}{2}\right)
	\eeq
	is expressed in terms of annihilation and creation operators, $\ha a$ and $\ha a^\dagger$, and has eigenenergies $E_n = \hbar \omega (n+1/2)$. This system represents one of the simplest integrable models. 
	The analytically continued partition function, $Z_{\beta + it} = \Big(2 \sinh[(\beta \hbar + i t) \omega/2] \Big)^{-1}$, gives the SFF as 
	\beq \label{S_HO}
	S_{\beta, t}^\textsc{ho}(\omega)=\frac{\cosh (\beta \hbar \omega) - 1}{\cosh(\beta \hbar \omega) - \cos(\omega t)}. 
	\eeq
The system energies have a constant spacing, so the SFF, shown in Fig. \ref{fig:SFF-integrable}(a), is a periodic  function---of period $2 \pi/\omega$. As the system temperature is increased, the SFF  minimum, equal  to $\tanh^2 (\beta \hbar \omega/2)$, decreases. We verify that $e^{-\eta t}$  constitutes a good approximation around $t_0$ to characterize the  decay of the SFF after the initial Gaussian decay.

		\begin{figure}[h]
		\centering
		\includegraphics[width=0.85\columnwidth]{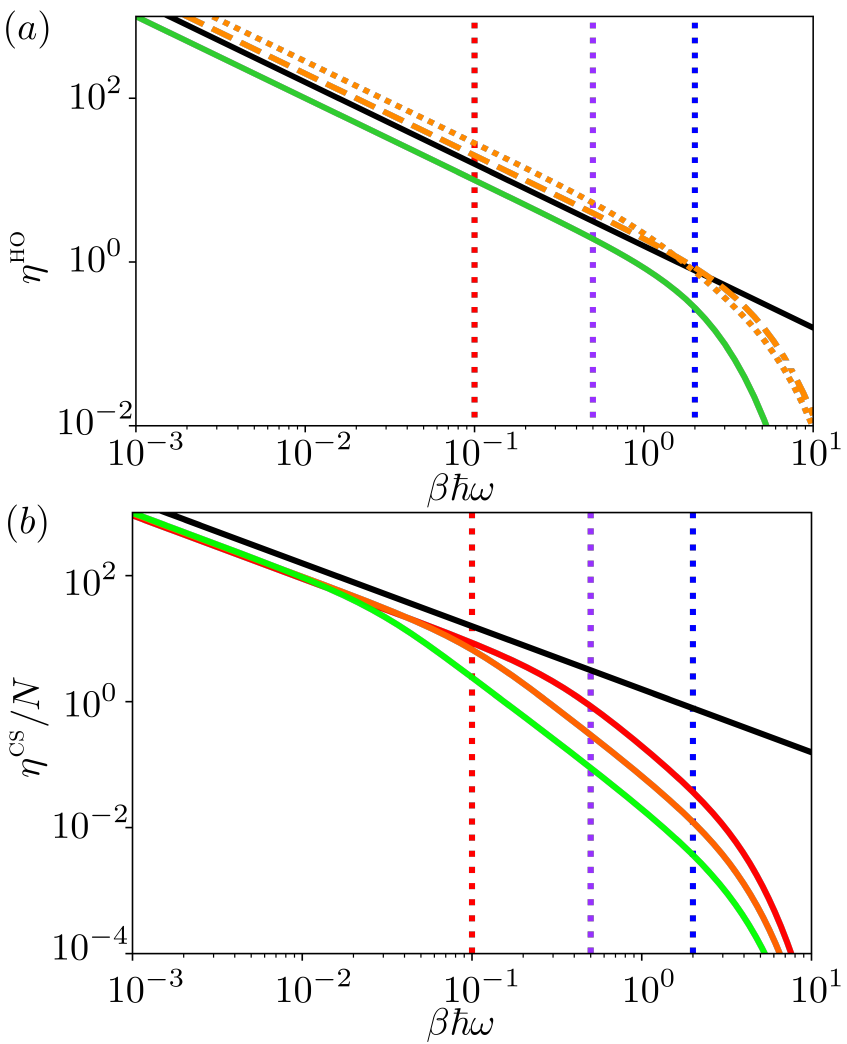}
		\caption{\textbf{Inflection exponent} and its bound \eqref{bound_eta} (black line) as function of   the dimension-less inverse temperature $\beta \hbar \omega$ for the (a)  harmonic oscillator $\eta$ (green) and (b)  Calogero-Sutherland model $\eta^\textsc{cs}_\beta/N$ for $N=10$ (red), $30$ (dashed orange) and $100$ (green). The vertical dotted lines represent the inverse temperatures shown in Fig. \ref{fig:SFF-integrable}. (a) The orange lines show bounds from QSL  $\eta_\textsc{qsl}$ \eqref{etaQSL} (dotted line) and from the Bhattacharyya results $\eta_\textsc{b}$ \eqref{etaB} (dashed line).}
		\label{fig:eta-integrable}
	\end{figure}

	In order to  obtain the inflection exponent $\eta$, defined in Eq. (\ref{def_eta}), we look for the time $t_0$ at which $\frac{\dot{S}}{S}=-\frac{\omega \sin (\omega t)}{\cosh(\beta \hbar \omega) - \cos (\omega t)}$ has an extremum. 
	%This imposes $\sin^2 (\omega t_0)= \cos(\omega t_0)\big[\cosh(\beta \hbar \omega)- \cos (\omega t_0)\big]$, that is, $\cos (\omega t_0)= 1/\cosh(\beta \hbar \omega)$. Using $1=\sin^2 x + \cos^2 x =\cosh^2 x - \sinh^2 x$, we get $\sin (\omega t_0)= \tanh(\beta \hbar \omega)$. 
	This gives the exponent as  
	\beq \label{eta_HO}
	\eta^{\textsc{ho}}_\beta(\omega) = \frac{\omega}{\sinh (\beta \hbar \omega)} =2 \,\omega  Z_{2\beta}\, .
	\eeq 
This inflection exponent gets closer to the $\pi/(2 \hbar \beta)$ bound (\ref{bound_eta}) at high temperature, with an asymptote at   $1/\hbar \beta$, as illustrated in Fig. \ref{fig:eta-integrable}(a).

	\subsection{Many-body integrable system: the Calogero-Sutherland model \label{sec:CS}}
	
	The Calogero-Sutherland (CS) model is a many-body system of $N$ particles in one dimension with inverse-square interactions \cite{calogero_solution_2003, sutherland_quantum_1971} that gives  insight into black-hole physics \cite{claus_black_1998,gibbons_black_1999,lechtenfeld_calogero_2016}. The Hamiltonian reads, in first quantization, 
	\begin{equation}
		\hat{H} = \sum_{n = 1}^{N} \left( -\frac{1}{2}\frac{\partial ^2}{\partial x_n^2}+ \frac{1}{2} \omega^2  x^2_n\right)+\sum_{n<n'}\frac{\ell (\ell-1)}{(x_n - x_{n'})^2},
	\end{equation}
	where $\ell$ determines the interaction strength and $\omega$ is the frequency of the harmonic trap in which the particles are confined. This model is equivalent to an ideal gas of Haldane anyons \cite{haldane_fractional_1991, wu_statistical_1994}, so its partition function  factorizes---as expected for an ideal gas \cite{murthy_thermodynamics_1994}. This factorization leads to the SFF being the product of the SFF for $N$ harmonic oscillators \eqref{S_HO} with increasing frequencies $n \omega$, namely \cite{jaramillo_quantum_2016, del_campo_scrambling_2017}
	\begin{equation} \label{S_CS}
		S_{\beta, t}^\textsc{cs} = \prod_{n = 1}^{N} \frac{\cosh(n \hbar \omega \beta)-1}{\cosh(n \hbar \omega \beta)- \cos(n \omega t)}.
	\end{equation}
	Note that this simple result does not hold when the trap frequency is time dependent \cite{campo_exact_2016}. The behavior of this SFF is illustrated in Fig. \ref{fig:SFF-integrable}(b): it is also $2 \pi/\omega$ periodic  but shows higher order peaks corresponding to the `harmonics' $n \omega$, particularly apparent for $n = 2$ and $n = 3$. Increasing the temperature $1/\beta$ or the number of particles $N$ (see inset) makes the SFF reach lower values.

	The inflection exponent  easily follows from $\ln S_{\beta, t}^\textsc{cs}= \sum_n \ln S_{\beta, t}^\textsc{ho}(n \omega)$ as 
	\begin{equation} \label{etaCS}
		\eta^\textsc{cs}_\beta = \sum_n \eta^\textsc{ho}_{\beta}(n \omega)=\sum_{n = 1}^{N} \frac{n \omega}{\sinh (n \beta \hbar \omega)}.
	\end{equation}
Its dependence with the inverse temperature is shown in Fig. \ref{fig:eta-integrable}(b). To compare it with the bound, we use the intensive quantity $\eta^\textsc{cs}/N$, as conjectured in \eqref{etaD}. The many-body interaction  effectively leads to two main regimes: one at very high temperatures in which the behavior is similar to a single harmonic oscillator, \textit{i.e.} $\eta^\textsc{cs}/N \sim \beta^{-1}$, and one at intermediate temperatures in which  $\eta^\textsc{cs}/N$ decays faster. This region grows with the number of interacting particles $N$.

	\subsection{Chaotic dynamics: random matrix ensemble} We now look at a typical chaotic system chosen within the common playground of random matrix theory \cite{wigner_statistical_1951-1, wigner_results_1956, mehta_random_2004, cotler_black_2017, cotler_chaos_2017, del_campo_scrambling_2017, chenu_quantum_2018, chenu_work_2019}. A Hermitian system with independent matrix elements and no time-reversal symmetry is represented by the Gaussian Unitary Ensemble (GUE) \cite{mehta_random_2004}. Averaging over a random matrix ensemble yields eigenenergies which are correlated in the same way as in a quantum chaotic system, according to the \textit{Bohigas-Giannoni-Schmit conjecture} \cite{bohigas_characterization_1984, bohigas_spectral_1984}. 
	A constituent of the ensemble is constructed by sampling every matrix element from a Gaussian distribution with standard deviation $\sigma = \hbar \omega_\textsc{gue}$. For the diagonal elements the Gaussian is real and for the off-diagonal it is complex, therefore the constructed Hamiltonian will be Hermitian.
	The ensemble averaging of the SFF \eqref{SFF} should rigorously be taken such that 
	$\left \langle \frac{|Z_{\beta + i t/\hbar}|^2}{Z_\beta^2}\right \rangle$ to represent physically measurable quantities, but the  `annealed' version,  with the average split as $\frac{\langle|Z_{\beta + i t/\hbar}|^2\rangle}{\langle Z_\beta^2\rangle}$, is useful to obtain analytical results. 
	Both averages are equal in the high-temperature limit. In the context of random matrix theory, the ensemble averaged SFF is commonly split into three terms,
	\begin{equation} \label{Z_GUE}
	S^\textsc{gue}_{\beta, t}= \frac{\langle Z_{2 \beta}\rangle + |\langle Z_{\beta + i t/\hbar} \rangle|^2 + g_c(\beta, t)}{\langle Z_\beta \rangle ^2},
	\end{equation}
	where the connected SFF $g_c(\beta, t)$ is detailed in App. \ref{app:GUE}. The averaged partition function for the GUE in a $\mc N$-dimensional Hilbert space is known as  
	\cite{del_campo_scrambling_2017} 
	\begin{equation} \label{avZGUE}
	\av{Z_{\beta + \frac{it}{\hbar}}}=e^{\frac{(\beta+ it/\hbar)^2}{4}} L^1_{\mc N{-}1} \Bigg({-}\frac{(\beta{+}it/\hbar)^2}{2}\Bigg),\hspace{-0.5em}
	\end{equation}
	where $L_n^\alpha (x) = \sum_{j=0}^n \binom{n+\alpha}{n-j} \frac{(-x)^j}{j!} $ are the generalized Laguerre polynomials.

\begin{figure}[h]
	\centering
		\includegraphics[width=0.85\columnwidth]{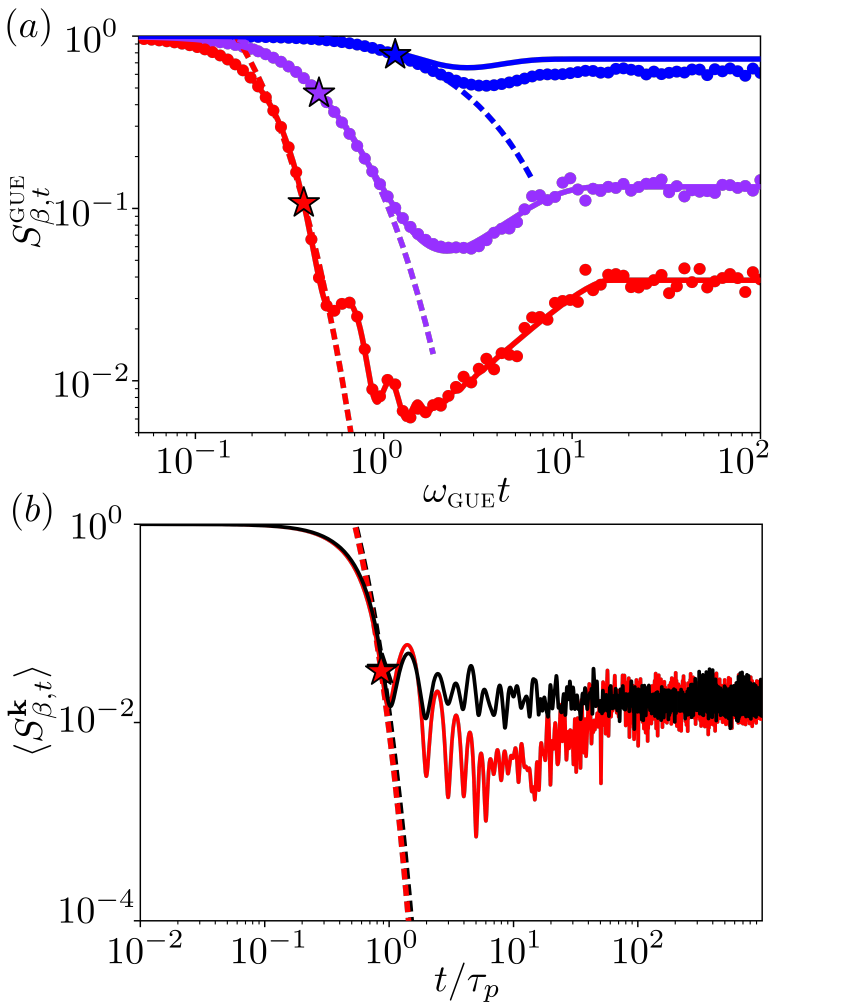}
	%	\vspace{-2ex}
		\caption{\textbf{Time evolution of the SFF} for  (a) the Gaussian Unitary Ensemble of dimension $\mc N=30$ and (b) the  quantum kicked top. The dashed lines represent the function $e^{-\eta t}$ around the inflection time $t_0$, marked by a star.  (a)  The dots represent the numerical average over $N_\mr{av}=100$ realizations, the solid lines represent the `annealed' analytical expression (\ref{Z_GUE}) for  $\hbar \beta  =2$ (blue), $0.5$ (purple) and $0.1$ (red).  (b) SFF for the quantum kicked top (\ref{avpseudoSFF}) in  the regular (black) and chaotic (red) regime, at  $\hbar \beta =0.1$, with spin $S=30$ and a numerical average over $N_\mr{av}=30$ realizations. \label{fig:SFF-chaotic}}
	%	\vspace{-2ex}
	\end{figure}

	Figure \ref{fig:SFF-chaotic}(a) shows the SFF computed numerically and analytically for the GUE. The behavior displays the shape (slope-dip-ramp-plateau) characteristic of chaotic systems. As the system temperature is decreased, the dip becomes shallower and occurs later. This is because the SFF accounts for all the possible energy correlations across the full spectrum: as the temperature is lowered, the contributions from neighbors further apart in energy---that have a smaller dip time---decreases, such that the dip time is delayed. This behavior is explicit from an expression of the SFF as function of the energy neighbors that we give in App. \ref{app:GUE}. %We note that the oscillations present at very high temperatures vanish at lower temperatures. 
	
	The function $e^{-\eta t}$ around the inflection point is also shown in Fig. \ref{fig:SFF-chaotic}(a). The dependence of the $\eta$ exponent as a function of the inverse system temperature is shown in Fig. \ref{fig:eta-chaotic}(a), together with its bound. We see that the exponent  gets close to the bound \eqref{bound_eta} for $0.1 \lesssim \beta \hbar \omega_\textsc{gue} \lesssim 1$. Interestingly, the exponent saturates to a constant value at high temperature, a feature not present in the harmonic oscillator, that is related to the finiteness of the Hilbert space $\mc N$: beyond some high enough temperature, all energy levels are already included within the thermal average and the saturation happens.

	\subsection{Tunable dynamics: generalized quantum kicked top} We now look at a system which dynamics can be tuned from regular to chaotic motion, the quantum kicked top, which was designed in the early days of quantum chaos studies and remains an important playground \cite{haake_classical_1987, kus_symmetry_1987, scharf_kramers_1988, haake_kicked_1988, fox_chaos_1994, chaudhury_quantum_2009, haake_quantum_2010, yin_quantum_2021}. 
	Kicked tops model a spin $S$ system subject to a free precession and some  $\tau_p$ periodic kicks, the strength of which allows going from periodic orbits to chaotic dynamics. 
	The stroboscopic description of such a periodic system  is well characterized in terms of the Floquet operator, which captures the time evolution of the system over one period.

%	Tops can be designed so as to represent any members of a universality class, and 
We use the Floquet operator for the general unitary class introduced by Haake \cite{haake_quantum_2010}
	\begin{align}
	\ha{\mc U} =& \:e^{- i (\frac{p_z}{\hbar} \ha S_z + \frac{1}{(2 S + 1)\hbar^2}k_z \ha S_z^2)} \:e^{- i (\frac{p_y}{\hbar} \ha S_y + \frac{1}{(2 S + 1)\hbar^2}k_y \ha S_y^2)} \nonumber \\
	&\times e^{- i (\frac{p_x}{\hbar} \ha S_x + \frac{1}{(2 S + 1)\hbar^2}k_x \ha S_x^2)},
	\end{align}
	that can 	mimic the behavior of any members of a universality class displayed by random matrix theory according to the choice of parameters $\be p = (p_x, p_y, p_z)$ and $\be k = (k_x, k_y, k_z)$, where  $\ha{\be S}= (\ha S_x, \ha S_y, \ha S_z)$ are the general spin operators. For example, for $\be k\times \be p=0$, e.g. the only non-zero terms are $k_z=1$ and $p_z=10$, the system is integrable (the level spacings follow Poisson statistics) because of the extra symmetry $[\ha{\mc U}, \ha S_z]=0$ that brings an extra conserved quantity---the $z$-component of the angular momentum. There are choices of parameters that break time-reversal symmetry and therefore the system behaves similarly to the GUE, e.g. $\be p =(1.1, 1, 1)$ and $\be k = (4, 0, 10)$.

		The eigenvalues of the Floquet operator, $\ha{\mc U}\ket{\chi_j}= e^{- i \omega_j^\mathbf{k} \tau_p} \ket{\chi_j}$, allow defining the pseudo-frequencies $\omega_j^{\be k}$  \cite{wang_butterfly_2009-1, wang_generating_2010}. 	
	For our purpose, we use these pseudo-frequencies to define the pseudo-SFF as
	\beq
	S^{\be k}_{\beta, t} = \frac{\sum_{m,n}e^{-(\beta \hbar + i t) \omega_m^{\be k}}e^{-(\beta \hbar - i t) \omega_n^{\be k}}}{\left( \sum_m e^{-\beta \hbar \omega_m^{\be k}}\right)^2}.
	\eeq
	
\begin{figure}[h]
	\centering
		\includegraphics[width=0.85\columnwidth]{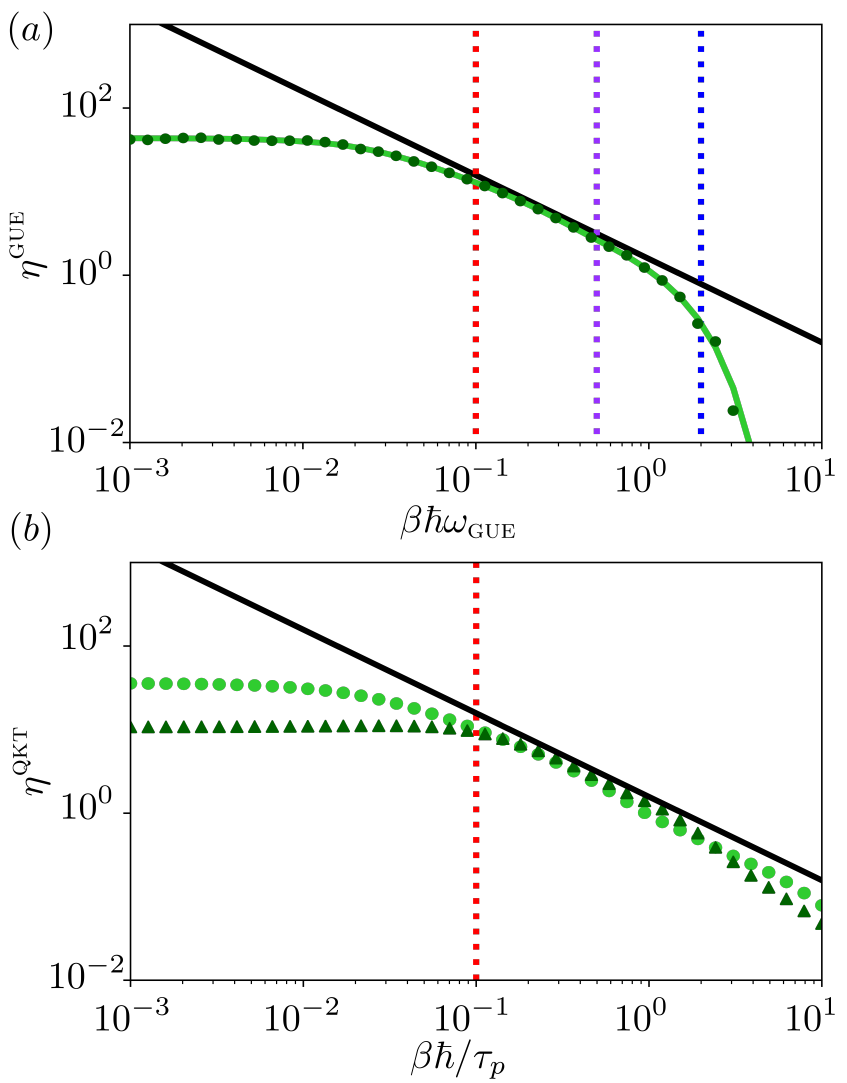}
		%\vspace{-2ex}
		\caption{\textbf{Inflection exponent $\eta$} (green) and its bound (Eq. \eqref{bound_eta}, black line) as function of the inverse temperature $\beta$ for  (a) the Gaussian Unitary Ensemble and (b) the quantum kicked top. The vertical dotted lines represent the inverse temperatures shown in Fig. \ref{fig:SFF-chaotic}.  Results in the quantum kicked top (b) are shown for the two dynamical regimes illustrated in Fig. \ref{fig:SFF-chaotic}(b), chaotic (green circles) and regular (dark green triangles). The inflection exponent $\eta^\textsc{qkt}$ gets close to the analyticity bound for $0.2 \lesssim \beta \hbar \lesssim 2$ in both dynamical regimes, which further confirms the independence of $\eta$ on the dynamics of the system. }
		\label{fig:eta-chaotic}
		%\vspace{-10ex}
	\end{figure}

	The SFF is in general not a self-averaging quantity \cite{prange_spectral_1997}, which means its behavior over one system realization generally differs from the ensemble average. To obtain an average behavior, we follow Haake's original idea \cite{haake_classical_1987} and introduce an averaging over some window of parameters. We uniformly generate $N_\mr{av}$ random points  in the interval $\mc K  \equiv (k_z-\delta k_z/2, k_z + \delta k_z/2)$ and average over them to obtain
	\beq \label{avpseudoSFF}
	\av{S_{\beta, t}^{\be k}}= \frac{1}{N_\mr{av}}\sum_{\kappa \in \mc K} S^{(k_x, k_y, \kappa)}_{\beta, t},
	\eeq
	where we choose $\delta k_z =0.05 k_z$. Fig. \ref{fig:SFF-chaotic}(b) shows the pseudo-SFF computed for the kicked top in the two dynamical regimes, regular and chaotic. In the latter, $\langle S_{\beta, t}^{\be k}\rangle$ exhibits the expected behavior in the chaotic phase, with a dip and a ramp at long times, absent in the former. Around the $t_0$ inflection point, both regimes behave quite similarly. This holds over a wide range of temperatures, as illustrated by the inflection exponent $\eta$ behavior in Fig. \ref{fig:eta-chaotic}(b). In both regimes, the exponent gets very close to the bound \eqref{bound_eta} imposed by analyticity, therefore the tightness of the bound is not related to regular or chaotic dynamics. The saturation at high temperatures  observed in the GUE is also present here because the kicked top has a finite dimensional Hilbert space, with $\mc N = 2 S +1$. 
	
%	\clearpage

	%%%%%%%%%%%%%%%%%%%%%%%%%
{\pablo 	
	\section{Relation to quantum speed limits and other known bounds \label{sec:QSL}} 
	We explore below the relation of the derived bound with other known bounds. 
}
	
	 Quantum Speed Limits (QSL) set a bound on the evolution of the fidelity.  For a pure state under unitary dynamics, the rate of change of the fidelity  $F_t = |\langle \psi_t | \psi_0 \rangle |^2$ is bounded by  \cite{shanahan_quantum_2018}
\begin{equation}\label{fidQSL}
	|\dot{F}_t| \leq \frac{\sqrt{2}}{\hbar} \Delta E,
\end{equation}
where $\Delta E =  \sqrt{\langle H^2\rangle -\langle H\rangle ^2}$ captures the energy fluctuations. Since the SFF is the fidelity of the pure, coherent Gibbs state, this bound applies to $S_{\beta,t}$ defined in Eq. \eqref{SFF}. 
In order to compare this bound with that on the inflection exponent \eqref{bound_eta}, we look at the inequality obtained from QSL at  time $t_0$, that yields
\begin{equation}
\eta =\frac{ |\dot{S}_{\beta,t_0}|}{S_{\beta,t_0}} \leq \eta_\textsc{qsl} \equiv \frac{\sqrt{2}}{\hbar} \frac{\Delta E}{ S_{\beta,t_0}}.
\end{equation}

For the example of the harmonic oscillator considered above, we easily get (see App. \ref{app:QSL} for details)
\begin{equation} \label{etaQSL}
\begin{split}
|\dot{S}_{\beta, t}^\textsc{ho}|&=\frac{|\omega \sin(\omega t) (1 - \cosh (\beta \hbar \omega))|}{(\cosh (\beta \hbar \omega) - \cos (\omega t))^2} \\
&\leq \frac{\sqrt{2}}{\hbar} \, \Delta E = \sqrt{2} \, \omega Z_\beta, 
\end{split}
\end{equation}
which further yields $\eta \leq \sqrt{2}\omega Z_\beta [\cosh (\beta \hbar \omega)+1]/ \cosh (\beta \hbar \omega)$. 
 Fig. \ref{fig:eta-integrable}(a) shows that the universal bound imposed by analyticity \eqref{bound_eta} is tighter than that imposed by QSL for temperatures above $\beta \hbar \omega \approx 2$. The asymptotic value of the QSL at  high temperatures is $2 \sqrt{2}/(\beta \hbar)$.

Also note that the survival probability, when larger than $1/2$, can be lower bounded by an exponential function, as shown by  Bhattacharyya \cite{bhattacharyya_quantum_1983}. This result has been extended to the spectral form factor  \cite{del_campo_scrambling_2017} and reads $S_{\beta,t} \geq e^{- 2 \Delta E t/\hbar}$. This lower bound gives an upper bound on the inflection exponent,  namely
\begin{equation}\label{etaB}
\eta \leq \eta_\textsc{b} = \frac{2}{\hbar} \Delta E. 
\end{equation}
For the harmonic oscillator, it is $\eta_\textsc{b}^\textsc{ho} = 2 \omega Z_\beta$. Fig. \ref{fig:eta-integrable}(a) compares all three bounds for the harmonic oscillator, in which system the universal bound set by analyticity constraints is the tightest at high enough temperature ($\beta \hbar \omega \lesssim 2 $). 

{\pablo 
 Bounds defined  from the temperature and the Planck constant only, so-called  Planckian bounds, have raised a renewed interest \cite{hartnoll2022, pappalardi2022, pappalardi_low_2022}. The bound on the Lyapunov exponent has thus been related to the fluctuation-dissipation theorem in the time domain  \cite{pappalardi2022}. Specifically, the OTOC corresponds to a two-point correlation function in a doubled Hilbert space, where the out-of-time ordering is introduced by a swap operator between the two Hilbert spaces. The two-point correlation function 
%\begin{align}
%S_\textsc{ab}(\beta, t) &= 
$\frac{1}{Z_\beta} \Tr(e^{-\beta \hat{H}} \ha A_t \ha B)$
%\end{align}	
can be split as $C_\textsc{ab}(t) + \hbar R''_\textsc{ab}(t) = \frac 1 {2 Z_\beta}\Tr(e^{-\beta \hat{H}} \{\ha A_t, \ha B\}) + \frac 1 {2 Z_\beta}\Tr(e^{-\beta \hat{H}} [\ha A_t, \ha B])$, that is, into terms  characterizing  fluctuations and  the response to external perturbations, respectively. 
The regulated form of this correlator, 
\begin{eqnarray} \label{Rcorrelator}
	\mc F_\textsc{ab}(\beta,t) = \frac 1 {Z_\beta} \Tr(e^{-\frac{\beta}{2} \hat{H}} \ha A_t e^{-\frac{\beta}{2} \hat{H}} \ha B ),
\end{eqnarray}
is similar to an OTOC when going to a doubled Hilbert space and introducing a swap operator.
By using the fluctuation-dissipation theorem, the authors in \cite{pappalardi2022} find a Planckian bound on two-point correlators, thus connecting with the MSS finding. Specifically, if the fluctuation term $C(t)$ decays exponentially, then $\mc F(t)$ decays exponentially as $\mc F(t) \sim e^{-t/\tau}$, with a rate bounded by
%\begin{equation}
$	\frac{1}{\tau} \leq \frac{\pi}{\beta \hbar}. $
%\end{equation}
This bound is actually also related to the one we derive, since, for $A_{m,n}= \frac{Z_\beta^{1/2}}{Z_{\beta/2}} \; \forall (m,n)$, the regulated two-point correlator corresponds to the SFF at double the temperature, 	$\mc F_\textsc{aa}(\beta,t) \equiv S_{\beta/2, t}$. We then have $\eta \leq \frac{\pi}{\beta \hbar}$, which matches our bound under the substitution $\beta \rightarrow \beta /2$ in \eqref{bound_eta}.
%, which is similar to the bound that we have found on the decay of the SFF considering double temperature. 
The main difference between the two approaches is that two-point correlation functions are not analytic at $t=0$ while we have shown that the SFF is analytic at $t=0$. %\begin{equation}
%	\eta \leq \frac{\pi}{\beta \hbar},

%It is also important to introduce the regulated 2-point correlation function as
%\begin{eqnarray}
%	F_\textsc{ab}(\beta,t) = \frac 1 {Z_\beta} \Tr(e^{-\beta H/2} \ha A_t e^{-\beta H/2} \ha B ),
%\end{eqnarray}
%and by choosing $\ha A$ to be a completely sparse operator, i.e. $A_{m,n}= \frac{Z_\beta^{1/2}}{Z_{\beta/2}} \; \forall m,n$ we find that the regulated 2-point correlator is the Spectral Form Factor at double the temperature
%\begin{eqnarray}
%	F_\textsc{aa}(\beta,t) \equiv S_{\beta/2, t}.
%\end{eqnarray}
%
%By using Fluctuation-Dissipation theorem the authors find that similar Planckian bounds apply on 2-point correlators. Particularly if $C(t)$ decays exponentially, the exponential decay of $F(t) \sim e^{-t/\tau}$ is bounded by
%\begin{equation}
%	\frac{1}{\tau} \leq \frac{\pi}{\beta \hbar},
%\end{equation}
%which is the same bound that we have found on the decay of the SFF considering double temperature
%\begin{equation}
%	\eta \leq \frac{\pi}{\beta \hbar},
%\end{equation}
%and therefore both bounds are related. The main difference between the two approaches is that 2-point correlation functions are not analytic at $t=0$ while we have shown that the SFF is analytic at $t=0$.

 The work \cite{grozdanov2021} finds bounds on transport coefficients through analyticity properties. The author uses univalence, the property of a complex function of being injective, and by finding domains in which the functions are univalent one can find bounds on physical properties like transport coefficients in hydrodynamic theories. 

}

{\pablo	
	\section{Conclusion} 
	We introduced the   \textit{inflection exponent} $\eta$ to characterize the early-time decay of the SFF, that happens after the initial Gaussian decay and we find can be approximated by an exponential. This exponent is related to the imaginary part of the average energy at complex temperature. Following arguments from complex analysis, we have found that it is bounded as $\eta \leq \pi/(2 \beta \hbar)$. 
%
%	The early-time decay of the SFF, just after the initial Gaussian decay, can be approximated to an exponential characterized by the \textit{inflection exponent} $\eta$, that we introduced. We have shown that, following arguments from complex analysis, this exponent  can be bounded as $\eta \leq \pi/(2 \beta \hbar)$. The inflection exponent is related to the imaginary part of the average energy at complex temperature.
		By contrast with the MSS bound on chaos, which is only saturated by black holes \cite{maldacena_bound_2016} and their holographic duals, like the Sachdev-Ye-Kitaev model \cite{kobrin_many-body_2021}, our bound on the SFF is already quite tight in a variety of  systems. 
	
}		
	We illustrated the bound in  systems representing regular and chaotic dynamics. At high temperature, the behavior of the exponent depends on whether the system Hilbert space is infinite dimensional or not. Indeed, this determines if more energy levels become available as the temperature increases, or not, in which later case the exponent saturates at a fixed value. Importantly, the behavior of $\eta$ is similar in the GUE and the quantum kicked top, even if the latter is tuned in the regular regime.

	Our results, based on analyticity constraints, set a  bound on the fidelity of the coherent Gibbs state. We show how they relate to known results from quantum speed limits, that set a bound on the fidelity based on unitary dynamics. Further investigation in this direction would look for possible extension 	of the bound set by the domain of analyticity to other dynamical quantities and even different domains of analyticity, which may change the functional dependence of the  quantities that can be bounded.
	
\medskip
\textbf{Acknowledgements} \par It is a pleasure to thank A. del Campo, J. Yang, A. Kundu, J. Pomar, S. Pappalardi, F. Roccati and R. Shir for insightful discussions and B. Mukhametzhanov and F. Balducci for comments on the manuscript. AC thanks the hospitality of the DIPC during revision of the manuscript. This work was partially funded by the Luxembourg National Research Fund (FNR, Attract grant  15382998) and by the John Templeton Foundation (Grant 62171). The opinions expressed in this publication are those of the authors and do not necessarily reflect the views of the John Templeton Foundation. 
	%This research was funded by the Luxembourg National Research Fund under an Attract project, QOMPET grant, 15382998. 

\appendix	
\setcounter{equation}{0}
\renewcommand{\theequation}{S\arabic{equation}}

{\pablo
\section{Analyticity of the Spectral Form Factor \label{app:AnalytSFF}}
We show below that the analytical continuation of the SFF, $S_{\beta , t+ i \tau}$, is analytic in the strip $-\beta \hbar \leq \tau \leq +\beta \hbar$ and $t\in \mbb R$.

A function of complex variable $f(x+i y) = u(x,y) + i v(x,y)$ is analytic at $z_0=x_0+ i y_0$ if and only if it is holomorphic, i.e. complex differentiable, at this point. For $f(z)$ to be holomorphic it has to obey the Cauchy-Riemann conditions at this point, 
$\partial_x u = \partial_y v$ and $\partial_y u = - \partial_x v$, 
and the partial derivatives $\partial_{x}u, \partial_y u, \partial_{x} v, \partial_y v$ have to be continuous at $z_0$. The SFF at complex time, defined in Eq. \eqref{analyticalSFF}, can be written as \begin{align} \notag
	S_{\beta, t + i \tau}&{=} \sum_m  \hspace{-0.2em}\frac{e^{-2 \beta E_m}}{Z^2_\beta} {+} \hspace{-0.4em} \sum_{m \neq n}   \hspace{-0.3em}\frac{e^{-\beta \bar E_{mn}+\tau\omega_{mn}}}{Z^2_\beta}\cos (\omega_{mn}t)\\ &-i\sum_{m \neq n} \frac{e^{-\beta \bar{E}_{mn}+\tau\omega_{mn}}}{Z^2_\beta}\sin (\omega_{mn}t),\\ \notag &\equiv u(t, \tau) + i v(t, \tau)
\end{align}
 where we introduced $\bar E_{mn}= E_m+E_n$ and $\omega_{mn}=(E_m - E_n)/\hbar$. 
The partial derivatives then read
\begin{align*}
	\partial_t u &= - \sum_{m \neq n} \frac{\omega_{mn}}{Z^2_\beta}e^{-\beta \bar E_{mn}+\tau\omega_{mn}} \sin (\omega_{mn}t) = \partial_\tau v,\\
%	\partial_\tau v &= - \sum_{m \neq n} \omega_{mn} e^{-\beta \bar E_{mn}+\tau\omega_{mn}}\sin (\omega_{mn}t) = \partial_t u \checkmark,\\
	\partial_\tau u &=  \sum_{m \neq n} \frac{\omega_{mn}}{Z^2_\beta} e^{-\beta \bar E_{mn}+\tau\omega_{mn}} \cos (\omega_{mn}t)=-\partial_t v .
%	\partial_t v &= - \sum_{m \neq n} \omega_{mn}e^{-\beta \bar E_{mn}+\tau\omega_{mn}}\cos (\omega_{mn}t)\\  &= -\partial_\tau u \checkmark,
\end{align*}
So the Cauchy-Riemann conditions are satisfied.

Let us now check if, for an infinite dimensional Hilbert space, the sums appearing in the definition of the SFF and its derivatives converge. 
To do so, we use the \textit{ratio test}: given a series of the form
$\sum_{n=0}^\infty a_n,$
the ratio test is based on the value of the limit 
$$L = \lim_{n \rightarrow \infty} \left|\frac{a_{n+1}}{a_n}\right|.$$
If $L <1$, the sum \textit{absolutely} converges, if $ L>1$ the sum diverges, and if $L=1$ the test is inconclusive. Recall that absolute convergence means $\sum_n|a_n|=L$. 

To check the convergence of the SFF
$$S_{\beta, t + i \tau}=\frac{1}{Z^2_\beta} \sum_{m=0}^\infty e^{-(\beta-\frac{\tau}{\hbar}+ i \frac{t}{\hbar})E_m}\sum_{n=0}^\infty e^{-(\beta+\frac{\tau}{\hbar}- i \frac{t}{\hbar})E_n},$$
we split the double sum into the product of two sums. This is true if at least one of the sum absolutely converges---\textit{Mertens' theorem} on the Cauchy product.
We thus look at the limit of each sum, that we denote $L_1$ and $L_2$. 
Considering that the energies are ordered $E_{m+1}\geq E_m$ and assuming no degeneracies at infinity, the test yields the following limits
\begin{align}L_1 %&= \lim_{m \rightarrow \infty } |e^{-(\beta-\frac{\tau}{\hbar}+ i \frac{t}{\hbar})(E_{m+1}-E_m)}|\\ \notag 
&= \lim_{m \rightarrow \infty } e^{-(\beta-\frac{\tau}{\hbar})(E_{m+1}-E_m)} <1 &\text{if }  \beta-\frac{\tau}{\hbar} >0,  \notag\\
L_2 %&= \lim_{n \rightarrow \infty } |e^{-(\beta+\frac{\tau}{\hbar}-i \frac{t}{\hbar})(E_{n+1}-E_n)}|\\ \notag 
& = \lim_{n \rightarrow \infty } e^{-(\beta+\frac{\tau}{\hbar})(E_{n+1}-E_n)} <1 &\text{if }  \beta+\frac{\tau}{\hbar} >0. \notag
\end{align}
So, the SFF converges within the region $-\beta < \frac{\tau}{\hbar} < \beta$. In addition, the `effective inverse temperature' $\beta \pm \frac{\tau}{\hbar}$ is positive and  the partition function also  converges. 

Now, to check that the sums in  the partial derivatives of the SFF also converge, we study the convergence of 
$$\sum_{m,n}(E_m-E_n)e^{-(\beta-\frac{\tau}{\hbar}+ i \frac{t}{\hbar})E_m}e^{-(\beta+\frac{\tau}{\hbar}- i \frac{t}{\hbar})E_n}. $$
We thus have  to check the convergence of two sums of the form
$$\sum_{m}E_me^{-(\beta\mp \frac{\tau}{\hbar}+ i \frac{t}{\hbar})E_m}\sum_n e^{-(\beta\pm \frac{\tau}{\hbar}- i \frac{t}{\hbar})E_n}. $$
The second sum was already shown to converge. The ratio test for the first sum gives
$$L = \lim_{m \rightarrow \infty } \left|\frac{E_{m+1}}{E_m}e^{-(\beta\mp \frac{\tau}{\hbar}+ i \frac{t}{\hbar})(E_{m+1}-E_m)}\right|.$$
Introducing the dimensionless level-spacing $s_m = \frac{E_{m+1}-E_m}{E_m}$, the limit can be rewritten as
$$L = \lim_{m \rightarrow \infty } (1+ s_m)e^{-(\beta\pm\frac{\tau}{\hbar})E_m s_m}.$$
Now, the function $f(x) = (1+x) e^{-a x}$ is smaller than one $\forall x>0$ if $a>1$. So the two series converge provided that
$$\big(\beta - \frac{\tau}{\hbar}\big)E_m>1 \; \rightarrow \tau <\beta \hbar -\lim_{m \rightarrow \infty }\left( \frac{\hbar}{E_m}\right),$$
$$\big(\beta + \frac{\tau}{\hbar}\big)E_m>1 \; \rightarrow \tau >-\beta \hbar +\lim_{m \rightarrow \infty }\left(\frac{\hbar}{E_m}\right).$$
Assuming that the spectrum is unbounded, \textit{i.e.} $E_m \rightarrow \infty$ for $m \rightarrow \infty$, we get the same strip $-\beta \hbar < \tau < \beta \hbar.$
% I think that asking for an unbounded spectrum is very physical, any density of states we can think of has tails that don't go to zero abruptly, but it is an extra condition. Even if this assumption is not well-justified the result seems to point out that we should make the strip smaller (and system dependent) but not that the SFF is not analytic in $t=0$. 
This verifies that the SFF is analytic in the strip $-\beta \hbar \leq \tau \leq +\beta \hbar$, including at $t=0$.

%\section{Extensivity of the bound and the modified strip}\label{app:Extensivity}

}
\section{Spectral form factor in the Gaussian Unitary Ensemble \label{app:GUE}}
\subsection{Connected SFF}
We first detail the connected SFF,
\beq
g_c(\beta, t){=}{\int} d E dE' \big\langle \rho_c^{(2)}(E, E')\big\rangle e^{-(\beta + \frac{i t}{\hbar}) E- (\beta - \frac{i t}{\hbar}) E'},
\eeq
 that  is  the double complex Fourier transform of the connected correlation function $\big\langle \rho_c^{(2)}(E, E')\big\rangle=\av{\rho(E) \rho(E')}-\av{\rho(E)}\av{\rho(E')}$, where $\av{\rho(E)}$ is the density of states and $\av{\rho(E) \rho(E')}$ is the 2-level correlation function which gives the probability density of finding a level around $E$ and another one around $E'$ \cite{mehta_random_2004}. An analytical expression is known for the GUE, and reads \cite{chenu_work_2019}
	\begin{align}
		g_c(\sigma, \sigma^*)=&- e^{\frac{\sigma^2 + {\sigma^*}^2}{4}}\sum_{n,m=0}^{\mc N-1} \frac{\min(m,n)!}{\Max(m,n)!}\times \\ \label{gcGUE}& \left( \frac{|\sigma|^2}{2} \right)^{|n-m|}\left| L^{|n-m|}_{\min(m,n)}\left( -\frac{\sigma^2}{2} \right) \right|^2, \nonumber
	\end{align}
	with the complex value $\sigma = \beta + \frac{i t}{\hbar}$. 
	
\subsection{Influence of the system size $\mc N$}
\begin{figure}
		\centering
		\includegraphics[width=0.85\linewidth]{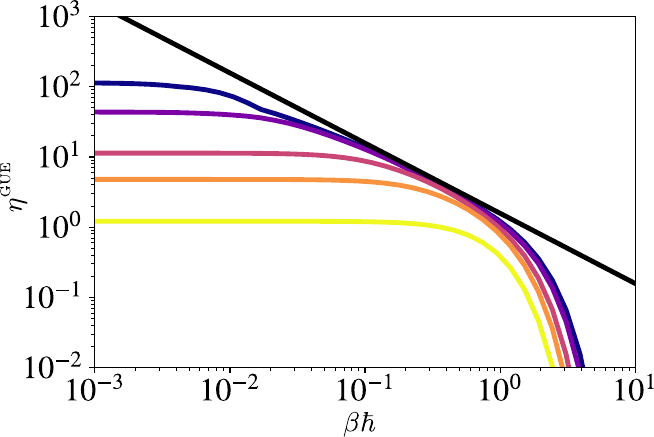}
		\caption{\textbf{Scaling of the inflection exponent} ${\eta}$ for the Gaussian Unitary Ensemble as a function of the inverse temperature $\beta$ for different system size: $\mc N=2$ (yellow), $5$ (orange), $10$ (pink), $30$ (purple) and $50$ (blue). Results are computed numerically from the analytical expression of the SFF for the GUE, 
		%Eq. (9) in the main text. 
		Eq. \eqref{Z_GUE}.
		 The black line represents the bound imposed by analyticity, 
		% Eq. (5) %
		Eq. \eqref{bound_eta} 
		in the main text. }
		\label{fig:scaling}
	\end{figure}

	Then, we look at the influence that the system size $\mc N$  has on the inflection exponent ${\eta}$.  Fig. \ref{fig:scaling} illustrates the role of large $\mc N$ in the region in which we get close to the bound imposed by analyticity,
	Eq. \eqref{bound_eta}  in the main text, 
%	Eq. (5) in the main text. 
	This region is observed to grow with the system size. Indeed for very low-dimensional Hilbert spaces, e.g. $\mc N=2$, the inflection exponent does not get close to the bound. The saturation value of the inflection exponent $\lim_{\beta \rightarrow 0} \eta$ at high temperatures is also seen to grow with the system size.
	
\subsection{SFF as function of the neighbor rank }

	From the definition of the SFF,
	\begin{equation}
	S_{\beta,t} = \frac{1}{Z_\beta^2}\sum_{n,m=1}^{\mc N} e^{- \beta (E_m + E_n)} e^{- \frac{i t}{\hbar} (E_n - E_m)},
	\end{equation}
	it is clear that this quantity carries information from all correlations across the full spectrum and not just those from nearest energy neighbors, which are  captured by the nearest-neighbor level spacing. These energy correlations are associated with chaotic behavior and give rise to the ramp, as discussed in the main text.

	The SFF may be written such as to make the role of the energy  correlations explicit. For this, we introduce the \textit{$j$-th level spacing} $s_n^{(j)}$  as the difference between $j$-th neighboring energies, namely $s^{(j)}_n=E_{n + j}-E_n$. The SFF becomes 
	\beq
		S_{\beta, t}= \frac{Z_{2\beta}}{Z^2_\beta} + \sum_{j=1}^{\mc N-1} S_{\beta, t}^{(j)},
	\eeq
	where $S^{(j)}_{\beta, t}$ is the contribution of the $j$-th energy neighbors, defined as
	\beq 
	\label{Sk_def}
		S^{(j)}_{\beta, t}=\frac{2}{Z^2_\beta} \sum_{n=0}^{\mc N-j}\cos \left(\frac{s^{(j)}_n t}{\hbar} \right) e^{-\beta \big(2 E_n + s_n^{(j)}\big)}.
	\eeq

\begin{figure}
		\includegraphics[width=1\linewidth]{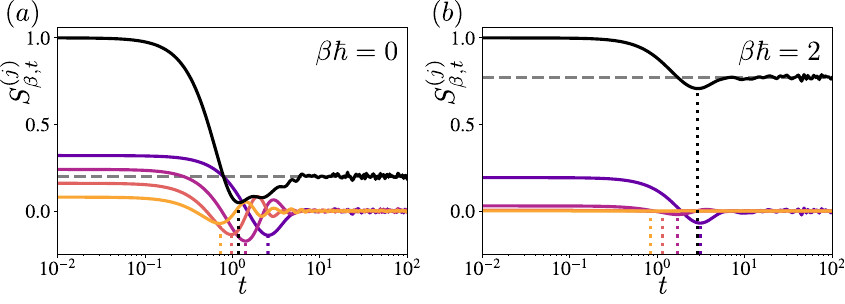}
		\caption{\textbf{Contributions from $\be j$-th neighbors $\boldsymbol{S^{(j)}_{\beta, t}}$ to the spectral form factor $\boldsymbol{S_{\beta,t}}$} (exact average) for the Gaussian Unitary Ensemble computed numerically at two different inverse temperatures (a) $\beta \hbar =0$ and (b) $\beta \hbar =2$. The full SFF $S_{\beta, t}$ (solid black line), reaches the plateau $\av{Z_{2 \beta}/Z_\beta^2}$ (grey dashed line) at large time. 
		The contributions $S^{(j)}_{\beta, t}$, as defined in \eqref{Sk_def}, are shown for $j=1$ (purple), $2$ (pink), $3$ (red) and $4$ (orange). The dotted lines mark the dip time of each contribution $S^{(j)}_{\beta, t}$. Here, $\mc N=5$ for clarity of the plot and the results are averaged over $N_\mr{av}=300$ realizations of the GUE.}
		\label{fig:RMT_Sk}
	\end{figure}

	Figure \ref{fig:RMT_Sk} shows the contributions of the different neighbor rank $j$ to the SFF. We see how the further away the energies are, that is, the larger the rank $j$, the sooner the dip time. This behavior is not surprising because for larger energy difference $s_n^{(j)}$, the time required to explore the full Hilbert space is shorter. At infinite temperature, Fig.~\ref{fig:RMT_Sk}(a) shows that the contribution at short times is larger for neighbors of lower rank, i.e. energies closer together.  The role of finite temperature can be understood from Fig. \ref{fig:RMT_Sk}(b), where the contributions for neighbors further apart, i.e. larger $j$, vanish with the term $e^{-\beta s^{(j)}_n}$ in \eqref{Sk_def}. This explains why the dip time is delayed  as the system temperature decreases, i.e. because the contributions for neighbors further apart in energy progressively vanish. This also shows how, at low temperatures, the SFF may be approximated from the contribution of nearest-neighbors $S^{(1)}_{\beta, t}$. This is reasonable since, as the temperature is lowered, the levels correlate less with levels further apart, and the most relevant contribution is captured by nearest-neighbors in energy.

\section{Quantum Speed Limits on the SFF for the Harmonic Oscillator \label{app:QSL}}

Quantum Speed Limits set a bound on the time derivative of the fidelity  of pure states given by \cite{shanahan_quantum_2018}
\begin{equation}\label{fidQSL}
	|\dot{F}_t| \leq \frac{\sqrt{2}}{\hbar} \Delta E.
\end{equation} 
The SFF may be interpreted as the fidelity between the coherent Gibbs state $\ket{\psi_\beta}$ and its time evolution, so the QSL on the fidelity yields a QSL on the SFF. The standard deviation of the energy thus needs to be taken with respect to the coherent Gibbs states, which mimic thermal averages, i.e. $ \braket{\psi_\beta}{\hat{H}^n |\psi_\beta}= \Tr (\hat{H}^n e^{-\beta \hat{H}})/Z_\beta= (-1)^n Z_\beta^{-1}\mr d^n Z_\beta/\mr d \beta^n$. The first two thermal moments, 
\begin{equation}
\begin{split} 
\langle \hat{H}\rangle &= -\frac{1}{Z_\beta}\frac{\mr d Z_\beta}{\mr d \beta}= \frac{\hbar \omega}{2}\coth \frac{\beta \hbar \omega}{2}, \\
\langle \hat{H}^2 \rangle &= \frac{1}{Z_\beta}\frac{\mr d^2 Z_\beta}{\mr d \beta^2}=\frac{(\hbar \omega)^2}{4}\left(2 \coth^2 \frac{\beta \hbar \omega}{2}-1\right),
\end{split}
\end{equation}
 yield the standard deviation of the energy $\Delta E = \sqrt{\langle H^2 \rangle -\langle H\rangle ^2}$ as
\beq \label{DeltaEHO}
\Delta E= \frac{\hbar \omega}{2}\sqrt{\coth^2 \frac{\beta \hbar \omega}{2}-1}=\frac{\hbar \omega}{2 \sinh \frac{\beta \hbar \omega}{2}}
\eeq
which simplifies to $\hbar\omega Z_\beta$. 

\begin{figure}[ht]
	\centering
	\includegraphics[width=1\linewidth]{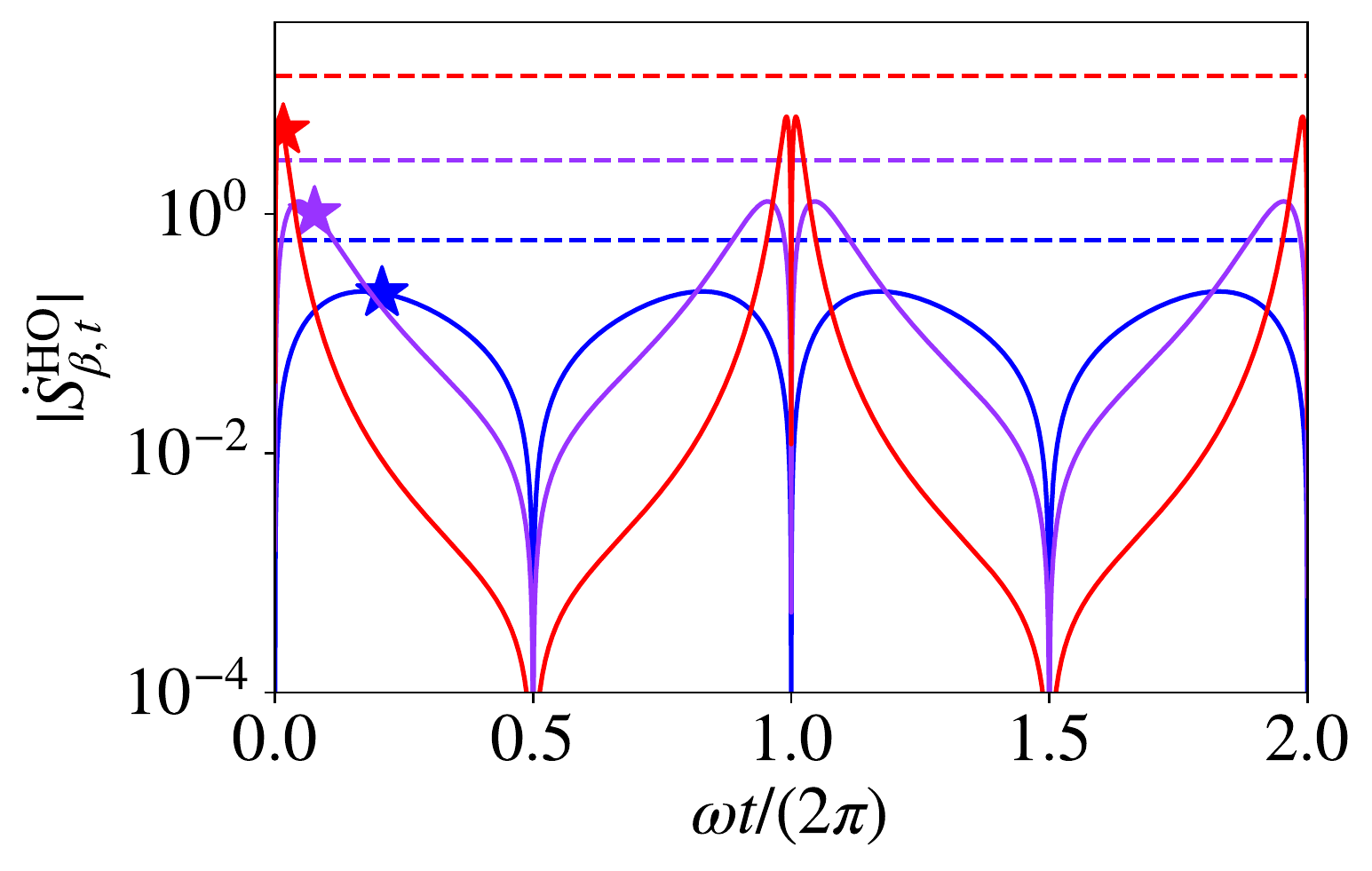}
	\caption{\textbf{Quantum Speed Limit} on the SFF for the harmonic oscillator at inverse temperatures $\beta =2$ (blue), $0.5$ (purple) and $0.1$ (red). The solid lines represent $|\dot{S}_{\beta, t}^\textsc{ho}|$, eq. (16) in the main text, bounded by  the QSL on fidelity \eqref{fidQSL}  (dashed lines), obtained with the energy standard deviation of Eq. \eqref{DeltaEHO}. The stars represent the inflection point of $\ln ( S_{\beta, t_0}^\textsc{ho})$. }
	\label{fig:HO_QSL}
\end{figure}

Figure \ref{fig:HO_QSL} shows the time derivative of the SFF together  with the bound set by the QSL on the fidelity \eqref{fidQSL}. This bound increases with the temperature, in a  fashion similar to the maximum value of $|\dot{S}^\textsc{ho}_{\beta, t}|$. The inflection point of $\ln( S_{\beta, t}^\textsc{ho})$ is close to the maximum of $|\dot{S}^\textsc{ho}_{\beta, t}|$ which corresponds to the inflection point of $ S_{\beta, t}^\textsc{ho}$.

%\bibliographystyle{apsrev4-1} % Tell bibtex which bibliography style to use
%\bibliography{TFM_Full_Biblio.bib,SFFbound_extra.bib,../../MEGA/MEGAlibrary/main,../../main}
%\bibliography{TFM_Full_Biblio.bib, SFFbound_extra.bib, main.bib, aux_pablo.bib}


%----------	
\end{document}